\documentstyle[12pt]{article}
\def\be{\begin{equation}}
\def\ee{\end{equation}}
\def\bea{\begin{eqnarray}}
\def\eea{\end{eqnarray}}

\begin{document}
\pagestyle{empty}
\begin{flushright}
{BROWN-HET-1054/TA-541}\\
{July 1996}
\end{flushright}
\vskip .25in
\begin{center}
MATRIX MODELS, OPEN STRINGS AND QUANTIZATION OF MEMBRANES\footnote{A talk given
at the {\it Argonne Duality Institute}, Argonne National Laboratory, June
27-July 12, 1996.}\\
\vskip .15in
by\\
\vskip .15in

Antal JEVICKI\\
{\it Department of Physics}\\
{\it Brown University}\\
{\it Providence, Rhode Island  02912  USA}\\
\vskip .25in
{\bf ABSTRACT}
\end{center}

We present an approach to membrane quantization using matrix quantum mechanics
at large $N$.  We show that this leads 
(through a simple field theory of two-dimensional open strings and the
associated $SU(\infty$) current algebra) to
a 4-D dynamics of self-dual gravity plus matter.
\vskip .10in

\newpage
\setcounter{page}{1}
\pagestyle{plain}

The problem of quantizing relativistic membranes (and higher dimensional
$p$-branes) is of major relevance.  These objects are seen to appear as
extended solitonic states of ten-dimensional superstring theory
\cite{one,two,three,four} and participate in the corresponding weak-strong
coupling duality maps.  Even more, the membrane is expected to play a
fundamental role as the basic object of the 11(12) dimensional $M$ (F) theory
\cite{five,six,seven} .

A fruitful idea for constructing the world volume description of extended
$p$-branes is given by the notion of world sheet - target space duality
\cite{eight,nine}.  In this,
a target space field theory of lower dimensional
branes is conjectured to give a {\bf world} {\bf volume} description of a
higher dimensional brane.  In particular, a quantum theory of a membrane 
could arise as a field theory of 2-d strings.

Matrix models have given certain insight into a field theory of lower
dimensional non-critical strings.  We can then contemplate an approach to
membrane quantization following the development of string equations and low
dimensional string field theory.  Indeed it was shown \cite{ten,eleven} that
a quantum membrane can be represented by a matrix model corresponding to
a dimensional reduction of an $SU(N)$ (super) Yang-Mills theory.  
In the large
$N$ limit $( N\rightarrow \infty )$ the $U(N)$ gauge group becomes a symmetry
of area preserving diffeomorphisms representing a symmetry of the membrane
Hilbert space.  The nontrivial feature in the 
quantization of the membrane is
connected with the treatment of this symmetry.

In what follows, we will describe an approach to this problem.  We will begin
with the matrix model formulation and develop a continuum field theory
operating in the invariant subspace.  The field theory involves a current
algebra of $SU(R)$ type representing open string fields which appear
as an intermediate construct \cite{twelve} in this approach.  The nontrivial
matrix dynamics is thereby represented in terms of conformal fields.  The
$R\rightarrow \infty $ limit is then argued to be the theory of a membrane.  
In particular,
we show that in this limit the dynamics is described by a 4-d {\bf self-dual}
gravitational field. The quantum Hamiltonian for this self-dual theory is
then specified by the $SU(\infty)$ current algebra model.

To begin, it is a result of Ref. [11] that a relativistic 
(super-) membrane in
the light cone gauge

$$
X^{\pm} = X^D \pm X^0
$$

\noindent
reduces to a Hamiltonian for the transverse degrees of freedom

$$
X_{i} \, \left( \sigma_1 , \sigma_2 \right)\, ,  \quad i = 1,2, 
\cdots, D-2; \qquad
\bar{\Theta} \, , \, \Theta^1 \left( \sigma_1 , \sigma_2 \right)
$$

\noindent taking the form

$$
H = \int d^2 \sigma \left( {1\over 2} \sum_{i=1}^{D-2} \, \dot{X}_i^2
 + {1\over
2} \sum_{i< j} \, \left\{ X_i , X_j \right\}^2 + {i\over 2} \, \bar{\Theta} \,
\gamma^i \left\{ X_i , \Theta \right\} \right)
$$

\noindent with the constraint

$$
G \left( \sigma_1 , \sigma_2 \right) = \sum_i^{D-2} \, \left\{ X_i , \dot{X}_i
\right\} + \bar{\Theta} \Theta
$$

\noindent representing area-preserving diffeomorphisms.  The bracket

$$
\left\{ A, B\right\} = \epsilon^{rs} \, \partial_r \, A\, \partial_s \, B
$$

\noindent corresponds to a large $N$ limit of the matrix commutator and the
correspondence with matrix notation is given through

$$
X_i \left( t, \sigma_1 , \sigma_2 \right) = \lim_{N\rightarrow\infty} \, M^{ab}
(t).
$$

\noindent The matrix Hamiltonian

$$
H = Tr \, \left\{ {1\over 2} \, \sum_{i=1}^{D-2} \,  \dot{M}_i (t)^2 + \Sigma
\left[ M_i , M_j \right]^2 + {i\over 2} \bar{\Theta} \, \gamma^i \left[ X_i ,
\Theta \right]  \right\}
$$

\noindent represents a dimensionally reduced SUSY Yang-Mills theory with only
a time dimension preserved.

Consider in what follows the case of a 4-dimensional membrane ($D-2=2)$,
concentrating on the bosonic coordinates (we will simply comment on the
supersymmetric extension at the appropriate place).  The problem that we have
is  that of a two-matrix system at $N\rightarrow \infty$. Using ideas of
collective field theory, we represent this in terms of {\bf conformal fields}
as follows:  The nontriviality of the problem lies in the Gauss Law constraint:

$$
G = \left[ M_1 , \dot{M}_1 \right] + \left[ M_2 , \dot{M}_2 \right] 
\approx 0.
$$

\noindent Gauge symmetry

$$
M_i (t) \rightarrow V^+ (t) \, M_i \, V(t)
$$

\noindent allows one to diagonalize one of the matrices, for example $M_1$:

$$
M_1 (t) = {\rm Diag} \,\, \left( \lambda_i (t) \right),
$$

\noindent and then

$$P_1 = \dot{M}_1 \rightarrow \, p_i + \left[ \dot{V} V^+ , \lambda \right],
$$
where
$$
p_i = \dot{\lambda}_i \qquad 0 = 1,2, \cdots, N.
$$

\noindent The Gauss law constraint is then solved for the angle variables
$\dot{V} V^+$ giving

$$
P_1 = p_i \, \delta_{ij} + {Q_{ij}\over (\lambda_i - \lambda_j )},
$$

\noindent with $Q_{ij} = [M_2 , \dot{M}_2 ]$ being the $SU(N)$ charge of the
second matrix.  We now introduce a {\bf reduction}:

$$
Q_{ij} = \sum_{a=1}^R \, \psi_i^+ \, (a) \,\,\psi_j (a),
$$

\noindent where $\psi_i (a)$ represent quark degrees of freedom with 
$a=1,\dots, R$
representing flavor indices.  We expect that in the limit $R\rightarrow
\infty$  the elements of the original matrix dynamics are recovered.  The
Hamiltonian problem at hand is then

$$
H = \sum_{i=1}^N \, {1\over 2} \, \dot{\lambda}_i^2 \, + {1\over 2} \, \sum_{i<
j} \,\,{(\psi^{+a} \psi^b)_i ( \psi^{+b} \psi^a )_j\over (\lambda_i - \lambda_j
)^2} \,\, + V.
$$

\noindent This is a form of the dynamical spin Calogero problem.  Here the
operators

$$
\left( \psi^{+a} \, \psi^b \right)_i
$$

\noindent create open strings with $a,b=1, \cdots, R$ representing 
the Chan-Paton
factors of the $SU(R)$ group.
This system has a continuum representation in terms of a field theory of open
strings (formulated in collaboration with J. Avan and J. Lee). 
The collective
fields are introduced

$$
\sum_i^N \, \delta \, \left( x - \lambda_i (t) \right) \rightarrow \alpha_+
(x,t) \, - \alpha_- (x,t) \, ,
$$

$$ \sum_i \left( \psi^{+a} \psi^b \right)_i \, \delta \left( x - \lambda_i
\right) \rightarrow \, J_+^{ab} \, \left( x,t) - J_-^{ab} \, (x,t)\right),
$$

\noindent giving variables of a $U(1)$ and $SU(R)$ 
affine current algebras.  For the
compact case one has $x \rightarrow z = e^{i\varphi}$, which corresponds to a
unitary matrix $U = e^{iM}$.  The continuum Hamiltonian describing the dynamics
is

\bea
H &= & \int dx \left\{ {1\over 6} \, \left( \alpha_+^3 - \alpha_-^3 \right) +
\alpha_+ \, T (J_+ ) - \alpha_- T (J_- ) \right\}\nonumber\\
&+& c \sum_{\epsilon = \pm} \, {J_{(x)}^{ab} \, J_{(y)}^{ba}\over (x-y)^2} +
\gamma^{\prime} \left( W_0 (J_+ ) - W_0 (J_- ) \right).
\eea

\noindent Here
$$T (J) = {1\over R + 1} \, \left( J^{ab} (x) \, J^{ba} (x) \right)
$$

\noindent is the energy-momentum tensor (of the current algebra) and $W_0$ are
the zero modes corresponding to  the spin-3 $W_3$ generators.

For details of the theory and a discussion of relevant symmetry properties of
the Hamiltonian the reader should consult refs. [12].

We have expressed the dynamics of the matrix model in terms of a $U(1)$ boson:
$$
\left[ \alpha_{\pm} (x) , \alpha_{\pm} (y) \right] = \pm 2 \, \partial_x \,
\delta (x-y)
$$

\noindent and  $L-R$  $SU(R)$  affine currents

\bea
\left[ J_{\pm}^{\alpha} (x) , J_{\pm}^{\beta} (y) \right] &= &
f_{\alpha\beta\gamma} J_{\pm}^{\gamma} \, \delta (x-y) \pm {\kappa\over 2\pi}
\, \delta^{\alpha\beta} \delta^{\prime}, \nonumber\\
\left[ J_+^{\alpha} , J_-^{\beta} \right]& = & 0,
\eea
representing degrees of freedom of a WZW model.
\noindent Let us now consider the limit $R\rightarrow \infty$ and exhibit the
associated continuum degrees of freedom that emerge.  In this limit, our
currents take the form

$$
J_{\pm}^{ab} (t,x) \rightarrow J_{\pm} \left( t,x,\sigma_1 , \sigma_2 
\right),
$$

\noindent thereby becoming a 4-dimensional field.  The scalar field
$\alpha_{\pm}$ obviously plays 
the role of a {\bf dilaton}.  Taking $\alpha_{\pm}
= \pm \alpha_0$, the quadratic term of the Hamiltonian reads

$$
H_2 = \int dx \, \alpha_0 \left( J_+^2 + J_-^2 \right),
$$

\noindent representing a Hamiltonian of an $SU(\infty)$ 
$\sigma$-model.  We now use
arguments due to Park \cite{thirteen} to show that the four-dimensional
dynamics that we have found is that of a self-dual gravity (the supersymmetric
extension of the theory can be found by simply extending the current
algebra).  Introducing

\bea
J_0 & = & J_+ + J_-, \nonumber\\
J_1 & = & J_+ - J_-, \nonumber
\eea

\noindent one has the algebra

\bea
\left[ J_0^{\alpha} , J_0^{\beta} \right] & = & f_{\alpha\beta\gamma} \,
J_0^{\gamma}, \nonumber\\
\left[ J_0^{\alpha} , J_1^{\beta} \right] & = & f_{\alpha\beta\gamma} \,
J_1^{\gamma} - \delta_{\alpha\beta} \delta^{\prime},\nonumber\\
\left[ J_1^{\alpha} , J_1^{\beta} \right]& = & 0, \nonumber
\eea

\noindent and the equations of motion

\be
\partial_t \, J_0 + \partial_x \, J_1 = 0,
\ee

\be
\partial_t \, J_1 - \partial_x \, J_0 + \left\{ J_0 , J_1 \right\} = 0,
\ee

\noindent where we have the Poisson bracket and the four-dimensional notation.
Solving the first equation with

\bea
J_0 & = & \partial_x \, \Omega \, \left( t, x, \sigma_1 , \sigma_2
\right),\nonumber\\
J_1 & = & - \partial_t \, \Omega \, \left( t, x, \sigma_1 , \sigma_2 \right),
\nonumber
\eea

\noindent one has for the second:

$$
\left( \partial_t^2 - \partial_x^2 \right) \, \Omega + \epsilon^{rs} \partial_r
\left( \partial_t \Omega \right) \partial_s \left( \partial_x \Omega \right) =
0.
$$

\noindent This equation of motion is associated \cite{fourteen,fifteen}
with the (Plebanski) action

$$
{\cal L} = \int dt \, dx \, d\sigma_1 \, d\sigma_2 \left( {1\over 2} \left(
\partial_t \Omega \right)^2 - {1\over 2} \left( \partial_x \Omega \right)^2 +
{1\over 3} \, \Omega \, \left\{ \partial_t \Omega , \partial_x \Omega \right\}
\right).
$$

We now make the following observation:  in the mechanism which produces
self-dual gravity, we recognize elements of non-abelian duality.  That is done
in the original works \cite{thirteen,fifteen} as a purely classical
transformation.  It is known that at the 
quantum level the correct procedure for
this is as in Refs. [16-19].  Consequently, the Plebanski theory is only
equivalent at the classical level and the full theory of quantized self-dual
gravity is to
be more precisely specified.

Let us then summarize the basic features of the present theory.  The
Hamiltonian

$$
H = \int \left( {\alpha^3\over 6} + \alpha \, J \left( x , \sigma
\right)^2\right)  + c \int \, {J (x, \sigma ) J (y, \sigma )\over (x-y)^2} +
\gamma \, W_0 (J)+V
$$

\noindent describes:

\begin{enumerate}
\item  a scalar field $\alpha$ representing a dilaton,

\item  a four-dimensional field $J (x,t,\sigma_1 , \sigma_2 )$ 
originating from the 
$SU(\infty )$ current algebra and representing a metric of 4-dimensional
self-dual gravity,

\item  an additional current-current interaction which we interpreted 
as coming
from matter (this explains the arbitrary coefficient $c$ present).

\end{enumerate}

We have found that, starting with the light cone quantum membrane, a 
four-dimensional world volume structure arises.  The precise form of
 4-dimensional
dilatonic self-dual gravity with matter is  based on the
{\bf nonabelian} dualization (with the corresponding Susy
extension) \cite{twenty}.  The quantum theory, however,  is defined by the
operator Hamiltonian presented in
terms of an $SU(\infty)$ Kac-Moody algebra.  
This then offers a framework for
studying the quantum spectrum.

\bigskip

\noindent{\large\bf Acknowledgement}
\medskip

\noindent It is a pleasure to thank Cosmas Zachos and Tom Curtright for their
hospitality and their organization of the Institute.

\end{document}